%% file: main.tex
\documentclass[sigconf]{acmart}
\AtBeginDocument{%
  }

\setcopyright{none}
\renewcommand\footnotetextcopyrightpermission[1]{}
\usepackage{algorithm}
\usepackage{algorithmic}
\usepackage{tabularx}
\usepackage{multirow}

\begin{document}

\title{Query-aware Routing for Filtered Approximate Nearest Neighbors Search}

 \author{Qianqian Xiong}
 \affiliation{%
   \institution{Australian National University}
   \city{Canberra}
   \country{Australia}}
 \email{secundussg@gmail.com}

\author{Mengxuan Zhang}
\authornote{Mengxuan Zhang is the corresponding author}
\affiliation{%
  \institution{Australian National University}
  \city{Canberra}
  \country{Australia}
}
\email{mengxuan.zhang@anu.edu.au}

\renewcommand{\shortauthors}{Xiong and Zhang}

\begin{abstract}
Filtered ANN search, which combines vector similarity with attribute predicates, is a core primitive in modern vector databases and retrieval-augmented generation. We benchmark all major categorical filtered ANN methods across multiple datasets under three predicates and find that no single method dominates. Moreover, even within a single dataset and predicate type, the best method for a query can vary.
Therefore, we propose a query-aware routing framework. A lightweight ML model predicts each candidate method's recall on the query, and the router consults an offline benchmark table that maps every method and parameter setting to its measured recall and QPS, then selects the method with the best recall--QPS trade-off. Our ablation study narrows 22 candidate features to a minimal set of three and we adopt regression rather than classification as the prediction target to sharpen accuracy. Our model is trained on six real-world datasets and applied to five unseen validation datasets. The final result shows that our router achieves state-of-the-art recall and QPS balance across all five validation datasets compared to existing filtered ANN baselines, while incurring negligible latency overhead.
\end{abstract}



\maketitle

\input{Introduction}
\input{Preliminary}
\input{MethodPart1}
\input{MethodPart2}
\input{RelatedWork}
\input{Experiment}

\input{Conclusion}

\bibliographystyle{ACM-Reference-Format}
\bibliography{reference}

\end{document}

%% file: Introduction.tex
\section{Introduction}
\label{sec:intro}

The development of deep learning and large pre-trained models has made it standard to encode unstructured data such as text, image, audio and video as high-dimensional dense vectors. Approximate Nearest Neighbor (ANN) search over such vector dataset has become a foundational component of recommendation systems, semantic retrieval, and Retrieval-Augmented Generation (RAG)~\cite{HNSW2018,DiskANN2019}. Given a query $q$ on a vector dataset $D$, ANN search retrieves the top-$k$ most similar vectors to $q$ from $D$. In real-world deployments, however, queries rarely involve only vector similarity. They typically carry additional attribute constraints, for instance, an e-commerce shopper searching for ``red dress'' requires both color and category to match; document retrieval could limit results by both time range and topic; Malt (Europe's largest freelancer marketplace) searches nearest neighbors with geo-spatial filtering.
This class of attribute-filtered approximate nearest neighbor search, known as \emph{filtered ANN}, has become a core capability of modern vector database systems such as Pinecone~\cite{Pinecone2024}, Milvus~\cite{Milvus2021}, Weaviate~\cite{Weaviate2024}, and hybrid analytical engines~\cite{AnalyticDB2020,VBASE2023}.

Filtered ANN queries are classified into two categories, depending on the filtered attribute type, namely \emph{range filtered ANN} filters by a numerical interval such as a timestamp or price band~\cite{SeRF2024,iRangeGraph2024,WindowFilters2024} and \emph{categorical filtered ANN} filters by a finite set of labels such as colors, categories, or tags. In this paper, we focus on the latter, whose solutions fall into two broad camps. The first camp consists of \emph{generic adaptations} on top of standard ANN indexes: Pre-filter~\cite{FilteredANNBenchmark2024} method filters candidates by label first and then selects the top-$k$ results over the filtered set, which guarantees recall but degenerates to linear scan when the filtered set is large; Post-filter~\cite{FilteredANNBenchmark2024} method first retrieves a large candidate pool with the top-$k'$ ($k'>k$) results by leveraging an ANN index and then filters them by verifying their labels, which is simple to deploy. But it requires $k'>>k$ to obtain $k$ valid results under high label selectivity (i.e., only a small fraction of vectors satisfy the predicate), which results in slow query processing. The second camp consists of \emph{specialized indexes} that use label information directly during index construction: UNG~\cite{UNG2024} indexes label-set containment relationships and supports cross-subgraph navigation; ACORN~\cite{ACORN2024} augments HNSW~\cite{HNSW2018} with expansion edges to preserve same-label reachability; FilteredVamana~\cite{FilteredStitchedVamana2023} integrates label-aware pruning into the Vamana graph~\cite{DiskANN2019}; SIEVE~\cite{SIEVE2025} pre-builds a workload-specialized collection of sub-indexes and picks the most suitable one per query. CAPS~\cite{CAPS2023} builds a two-level index with coarse $k$-means clusters refined by label-based sub-clusters; and NHQ~\cite{NHQNPG2023} builds a navigating proximity graph whose edge weights fuse vector distance with label Hamming distance.

\begin{figure}[t]
\centering
\includegraphics[width=\linewidth]{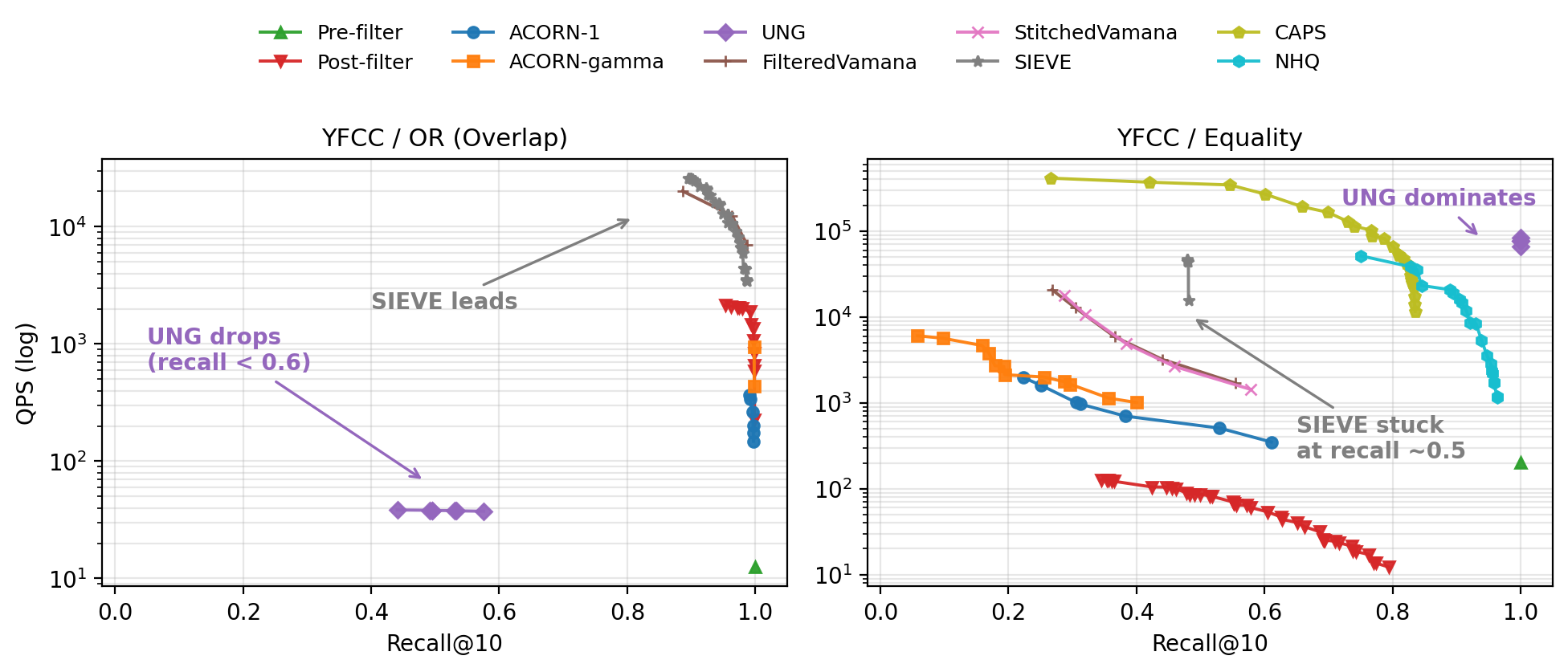}
\captionsetup{justification=centering, singlelinecheck=false}
\caption{Recall-QPS Pareto curves of filtered ANN methods on YFCC dataset under OR and Equality predicate types.}
\label{fig:per-scenario}
\end{figure}

\textbf{Observation.} Based on the predicate type, filtered ANN queries can be categorized into three types: \emph{Containment} (also known as AND), which requires results to satisfy all query labels; \emph{Overlap} (also known as OR), which requires results to satisfy at least one label; and \emph{Equality}, which requires the label set to match exactly. Through the preliminary experimental study, we observe that each method excels in its target setting, but no single method dominates across all combinations of datasets and predicate types (\underline{Observation 1}).
Figure~\ref{fig:per-scenario} contrasts representative filtered ANN methods on the YFCC dataset~\cite{Thomee2016YFCC100M} under two predicate types, i.e., OR and Equality. UNG and SIEVE exhibit opposite trends: UNG dominates Equality at recall = 1.0 but collapses on OR to recall $\approx$ 0.58, while SIEVE leads OR at recall > 0.9 but stalls on Equality at recall $\approx$ 0.5.
Moreover, even within the same dataset and predicate type, the best method varies from query to query (\underline{Observation 2}). For instance, we observe that some queries reach 100\% recall with UNG while Post-filter manages only 10\%, whereas others show the opposite pattern for AND queries on YFCC dataset. Therefore, given a filtered ANN query, it is essential to take the dataset, predicate type, and query into consideration to select the best-performing methods.

\textbf{Motivation.} Based on the above two observations, how about selecting the best-performing filtered ANN method each time a query comes, such that we can achieve the best query performance?
Motivated by Observation 1, we first attempted Rule-based Router (in Section~\ref{sec:method1}), i.e., assign one method based on both the dataset-level features and predicate type.
However, it was found that such a router can hardly route queries to their best-fit method, causing substantial recall loss. This happens because the query itself also determines the best-performed method according to Observation 2. We therefore propose to dive further by making routing decisions at the per-query level. Specifically, in the offline pre-processing stage, an ML model is trained to evaluate candidate methods for each query. And a lookup table is built by benchmarking each method across multiple parameter settings on every combination of dataset and predicate type, and record the corresponding query performance. Then, in the query processing stage, the router first builds a feature for one query and leverages the lookup table to select one combination of method and its parameter setting to achieve the best query performance.

\textbf{Challenges.} Nevertheless, it is non-trivial to design a router with both high-quality and fast decision making in selecting the best-performed method.

The first challenge lies in the \emph{Feature selection}. The key question is: how can we choose features that enable the ML model to accurately predict each candidate method's recall on a given query? To solve this question, we first explored over 20 per-query and dataset-level statistics, covering factors known to affect ANN search difficulty: geometric complexity (e.g., LID, relative contrast)~\cite{Amsaleg2015LID}, label structure (e.g., cardinality, entropy), label density, and label--vector coupling (e.g., distribution factor)~\cite{Lin2025FANNSSurvey,FilteredANNBenchmark2024}. However, using more than 20 features risks redundancy. Since additional features beyond the final chosen set add little predictive gain on the training set, while inflating the risk of overfitting training-specific patterns that fail to transfer to the validation set. Therefore, we use a nested feature ablation (Section~\ref{sec:experi}) to narrow these to a 3-feature minimal set: selectivity (per-query), LID (dataset-level geometric difficulty), and predicate type. 

The second challenge is how to design the \emph{model architecture}. The most intuitive choice is a classification model that directly outputs the best method for each query. However, the classification loss only sees whether the predicted method is correct, ignoring the actual recall gap, i.e., it does not distinguish the cost of being slightly off from being far off. We therefore switch to per-method regression with Mean Squared Error loss, which captures the magnitude of error, by assigning one independent MLP for each candidate method to predict the method's query performance.
Moreover, the router's latency is also critical since it directly affects the user-perceived query latency. That means, to achieve fast inference, the MLPs must remain lightweight (i.e., shallow with no extra vector computation) so as to produce negligible overhead to the per-query latency. 
Beyond accuracy, predicting recall alone is insufficient since different applications have different recall-QPS trade-offs. To handle this, the router uses an application-specified recall threshold $T$: at query time, it predicts each candidate method's recall, keeps only methods whose predicted recall meets $T$, and picks the parameter setting with the highest QPS from the offline lookup table. Since the model does not depend on $T$, a single trained model serves a flexible $T$.

\textbf{Contribution.} This paper makes the following contributions:

\begin{itemize}

\item We establish the empirical motivation of this work through an extensive experimental study over ten filtered ANN algorithms across six datasets, three predicate types, and thousands of parameter settings.

\item We identify a minimal feature set that matches the performance of larger feature sets while reducing variance and overfitting risk.

\item We propose a lightweight regression-based routing architecture with high recall.

\item We demonstrate that our proposed router generalizes robustly to unseen datasets, achieving near-optimal recall with negligible routing overhead.
\end{itemize}

%% file: Preliminary.tex
\section{Preliminaries}
\label{sec:preli}

We now formally define the filtered ANN problem and review its key evaluation metrics.

\subsection{Filtered ANN Problem Definition}
\label{ssec:problem}

Let the dataset $V = \{(v_i, L_i)\}_{i=1}^{n}$ contain $n$ vectors with attribute labels, where $v_i \in \mathbb{R}^d$ and $L_i \subseteq U$ is a discrete subset of a label universe $U$. A filtered ANN query is a triple $q = (x_q, L_q, P)$ comprising a query vector $x_q \in \mathbb{R}^d$, a query label set $L_q \subseteq U$, and a Boolean predicate $P : 2^U \times 2^U \to \{\mathrm{true}, \mathrm{false}\}$. Three predicate types are commonly used:
\begin{itemize}
    \item Equality: $P(L_i, L_q) = (L_i = L_q)$, requiring identical label sets.
    \item Containment (AND): $P(L_i, L_q) = (L_q \subseteq L_i)$, requiring the candidate to carry every query label.
    \item Overlap (OR): $P(L_i, L_q) = (L_q \cap L_i \neq \emptyset)$, requiring at least one shared label.
\end{itemize}

An entry $(v_i, L_i)$ is considered a valid candidate if and only if $P(L_i, L_q) = \mathrm{true}$; otherwise, it is excluded.
Given $q$ and an integer $k$, the goal is to retrieve $k$ vectors with the smallest Euclidean distance to $x_q$ from the valid candidates. ANN algorithms trade a small amount of accuracy for query speed and return an approximate top-$k$ set $R(q)$ as:
\begin{equation}
    R(q)=\mathop{\arg\min}\limits^{k}_{(v_i, L_i) \in V,\; P(L_i, L_q) = \mathrm{true}} \|v_i - x_q\|
\end{equation}

\subsection{Evaluation Metrics}
\label{ssec:metrics}

We adopt the two metrics standard in the ANN literature~\cite{ANNBenchmarks2020}:
\begin{equation}
    \mathrm{recall}@k(q) = \frac{|R(q) \cap \mathrm{TopK}(q)|}{\min(k,\; |\mathrm{TopK}(q)|)}, \quad
    \mathrm{QPS} = \frac{|Q|}{\sum_{q \in Q} t(q)}.
\end{equation}
$\mathrm{recall}@k$ measures the fraction of true top-$k$ retrieved by the algorithm; when fewer than $k$ candidates satisfy $P$, the denominator falls back to $|\mathrm{TopK}(q)|$ to avoid under-counting. $\mathrm{QPS}$ characterizes throughput as the test set size $|Q|$ divided by the total processing time, where $t(q)$ is the end-to-end latency of a single query.

%% file: MethodPart1.tex
\section{Rule-based Router}
\label{sec:method1}

\begin{figure*}[t]
    \centering
    \includegraphics[width=\linewidth,height=20cm,keepaspectratio]{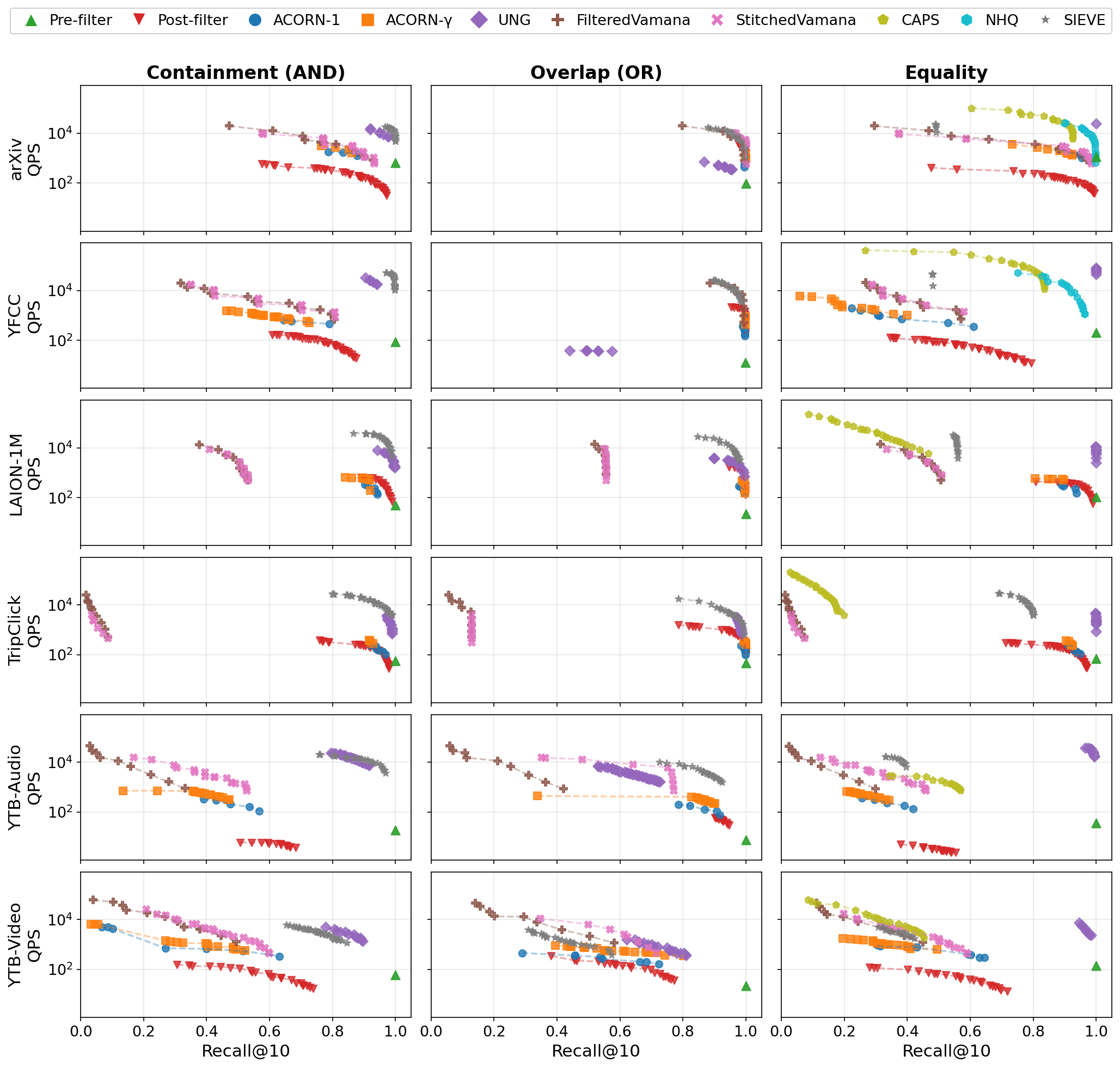}
    \caption{Recall--QPS comparison of all benchmarked filtered ANN methods across datasets and predicate types.}
    \label{fig:methods-comparison}
\end{figure*}

In this section, we introduce a rule-based router, motivated by a key observation from our empirical study: no single filtered ANN method dominates across all combinations of datasets and predicate types. Figure~\ref{fig:methods-comparison} shows the recall--QPS comparison of all benchmarked methods, where the best method shifts from one combination to another. A natural idea is to select a method with a small set of rules based on the dataset and predicate type. We instantiate this idea as a hand-crafted decision tree, which we call \underline{RuleRouter}.

For the dataset-level features, we use two characteristics of dataset\cite{Amsaleg2015LID} to demonstrate it:
\begin{itemize}
    \item \emph{Local Intrinsic Dimensionality (LID)}~\cite{Amsaleg2015LID} is a distance distribution based measure of local geometric complexity. For a query $q$ with its $k$ nearest-neighbor distances $r_1 \le r_2 \le \cdots \le r_k$, the maximum-likelihood per-query estimator is
\begin{equation}
    \mathrm{LID}(q) = -\left(\frac{1}{k} \sum_{i=1}^{k} \ln \frac{r_i}{r_k}\right)^{-1},
\end{equation}
and the dataset-level local intrinsic dimensionality $LID_{mean}$ is the average of $\mathrm{LID}(q)$ over a fixed sample $Q_{\mathrm{sample}}$ of queries:
\begin{equation}
    \mathrm{LID}_{\mathrm{mean}} = \frac{1}{|Q_{\mathrm{sample}}|} \sum_{q \in Q_{\mathrm{sample}}} \mathrm{LID}(q).
\end{equation}
Intuitively, high $\mathrm{LID}(q)$ indicates that many candidates lie at similar distances around $q$, making nearest-neighbor discrimination harder. So $\mathrm{LID}_{\mathrm{mean}}$ summarises this difficulty at dataset level.

\item \emph{Label cardinality.} For a dataset $V$ with label universe $U = \bigcup_i L_i$, the label cardinality is
\begin{equation}
    \mathrm{card}(V) = |U|.
\end{equation}
It is a coarse indicator of label-distribution sparsity: small $\mathrm{card}(V)$ means most queries share label sets with vectors in dataset, while large $\mathrm{card}(V)$ means each label is sparsely populated.
\end{itemize}

Therefore, we use both dataset-level features and predicate type as the decision features for RuleRouter: predicate type, $\mathrm{LID}_{\mathrm{mean}}$, and label cardinality.

\begin{table}[!htbp]
\centering
\captionsetup{justification=centering, singlelinecheck=false}
\caption{Best performed method for each combination of Dataset and Predicate type.
}
\label{tab:best-method-per-cell}
\small
\setlength{\tabcolsep}{4pt}
\begin{tabular}{@{}lrrccc@{}}
\toprule
Dataset & $\mathrm{LID}_{\mathrm{mean}}$ & $\mathrm{card}(V)$ & Equality & AND & OR \\
\midrule
arxiv      &  25.5 &   4{,}231 & UNG & SIEVE       & Post-filter \\
yfcc       &  23.0 & 181{,}931 & UNG & SIEVE       & Post-filter \\
LAION-1M   &  36.3 &        30 & UNG & UNG         & Post-filter \\
tripclick  &  31.5 &        29 & UNG & UNG         & Post-filter \\
ytb\_audio &  20.5 &   3{,}862 & UNG & SIEVE       & Post-filter \\
ytb\_video & 236.0 &   3{,}862 & UNG & UNG         & UNG         \\
\bottomrule
\end{tabular}
\end{table}

Then, we benchmark all methods on all combinations of training datasets and predicate types, with the best performed method in each combination shown in Table~\ref{tab:best-method-per-cell}.
As analyzed based on three decision features: (i) When the predicate type is Equality, UNG wins on every dataset. Since UNG's navigating graph clusters vectors with identical label sets, this is exactly what an exact-match predicate exploits. (ii) When predicate type is AND, the winner splits by label cardinality and $\mathrm{LID}_{\mathrm{mean}}$. When $\mathrm{card}(V)$ is small (LAION-1M, tripclick: only $\sim$30 distinct labels) or $\mathrm{LID}_{\mathrm{mean}}$ is high (ytb\_video, $\mathrm{LID}_{\mathrm{mean}}=236$), the filtered candidate pool is small or hard to search and UNG remains safest; otherwise SIEVE's pre-built sub-indexes deliver higher QPS. (iii) When predicate type is OR, the pattern follows $\mathrm{LID}_{\mathrm{mean}}$ alone. With high $\mathrm{LID}_{\mathrm{mean}}$ such as on ytb\_video, UNG's same-label connectivity beats brute candidate enumeration of Post-filter; on the remaining datasets with low-$\mathrm{LID}_{\mathrm{mean}}$, Post-filter on a generic ANN index suffices. Therefore, we formalize these patterns as the decision tree as shown in Algorithm~\ref{alg:rulerouter}.

\begin{algorithm}[!htbp]
\caption{RuleRouter: routing based on dataset-level features}
\label{alg:rulerouter}
\begin{algorithmic}[1]
\REQUIRE predicate type $pt$, dataset features $\mathrm{LID}_{\mathrm{mean}}$ and $\mathrm{card}(V)$
\ENSURE routing method $m$
\IF{$pt = \text{Equality}$}
    \STATE \textbf{return} UNG
\ELSIF{$pt = \text{AND}$}
    \IF{$\mathrm{LID}_{\mathrm{mean}} > 100$ \textbf{or} $\mathrm{card}(V) < 100$}
        \STATE \textbf{return} UNG
    \ELSE
        \STATE \textbf{return} SIEVE
    \ENDIF
\ELSIF{$pt = \text{OR}$}
    \IF{$\mathrm{LID}_{\mathrm{mean}} > 100$}
        \STATE \textbf{return} UNG
    \ELSE
        \STATE \textbf{return} Post-filter
    \ENDIF
\ENDIF
\end{algorithmic}
\end{algorithm}

Since RuleRouter operates based on the decision features related to dataset and predicate type, all queries within one combination of dataset and predicate type are routed to the same method. However, as revealed in Observation 2, for queries in the same combination of dataset and predicate type, their best performance methods are different. That means, the RuleRoute ignores the query-aware feature.
For example, on yfcc dataset with predicate type of AND, some queries achieve $100\%$ recall using UNG while Post-filter yields only $10\%$ recall, and other queries show the opposite pattern. This difference stems from query-aware features, such as the number of labels carried by query and per-label frequency statistics.%

To improve the routing quality, we enrich the feature set by including the query-aware features, and training an Machine Learning model (ML model) to select the best performed method.
However, this design faces two core challenges. The first is \emph{feature selection}, as the candidate feature space grows once we union query-aware and dataset-level features. What's worse, too many features would potentially result in overfitting with only six training datasets.
The second challenge is \emph{model architecture}, as the router needs to select the best performed method per query, it could be time-consuming. So there will be extremely high requirement on the balance of the model's inference speed and recall.

%% file: MethodPart2.tex
\section{ML Router}
\label{sec:method2}

To help understand our query-aware ML Routing system better, we first formalize the per-query routing setup. Let $\mathcal{M} = \{m_1, m_2, \ldots\}$ be a set of candidate filtered ANN methods. Each method $m \in \mathcal{M}$ exposes one or more tunable parameter settings, which we denote by $\mathit{ps} \in \mathcal{S}$, where $\mathit{ps}$ is the parameter setting of one method (e.g., \texttt{$L_\text{search}$} for UNG) and $\mathcal{S}$ is the union of all methods' parameter spaces. We use an offline benchmark table $B$ indexed by dataset $\mathit{ds}$, predicate type $\mathit{pt}$, method $m$, and parameter setting $\mathit{ps}$:
\begin{equation}
    B[\mathit{ds}, \mathit{pt}, m, \mathit{ps}] = \big( \mathrm{recall}_{\mathit{ds},\mathit{pt},m,\mathit{ps}},\; \mathrm{QPS}_{\mathit{ds},\mathit{pt},m,\mathit{ps}} \big)
\end{equation}

Table $B$ records the average performance of each method under each specific parameter setting, by testing the query performance under all combinations of predicate type and dataset.
A router $\mathcal{R}$ takes a query $q$, the dataset $\mathit{ds}$, and a specified recall threshold $T$, predicts the query performance of each method $m$, and then selects the best-performing method satisfying $T$ along with its parameter setting, i.e., $\mathcal{R}(q, \mathit{ds}, T) \to (m, \mathit{ps})$.

After that, the system will process the filtered ANN query by using method $m$ with parameter setting $\mathit{ps}$ to retrieve the top-$k$ results. We assume $B$ is obtained via offline benchmarking, which is standard practice for stable datasets (document collections, knowledge bases) and is routinely employed in industrial-scale ANN systems~\cite{Pinecone2024,Milvus2021}.

\subsection{Overview}
\label{ssec:overview}

\begin{figure*}[!t]
    \centering
    \includegraphics[width=\linewidth]{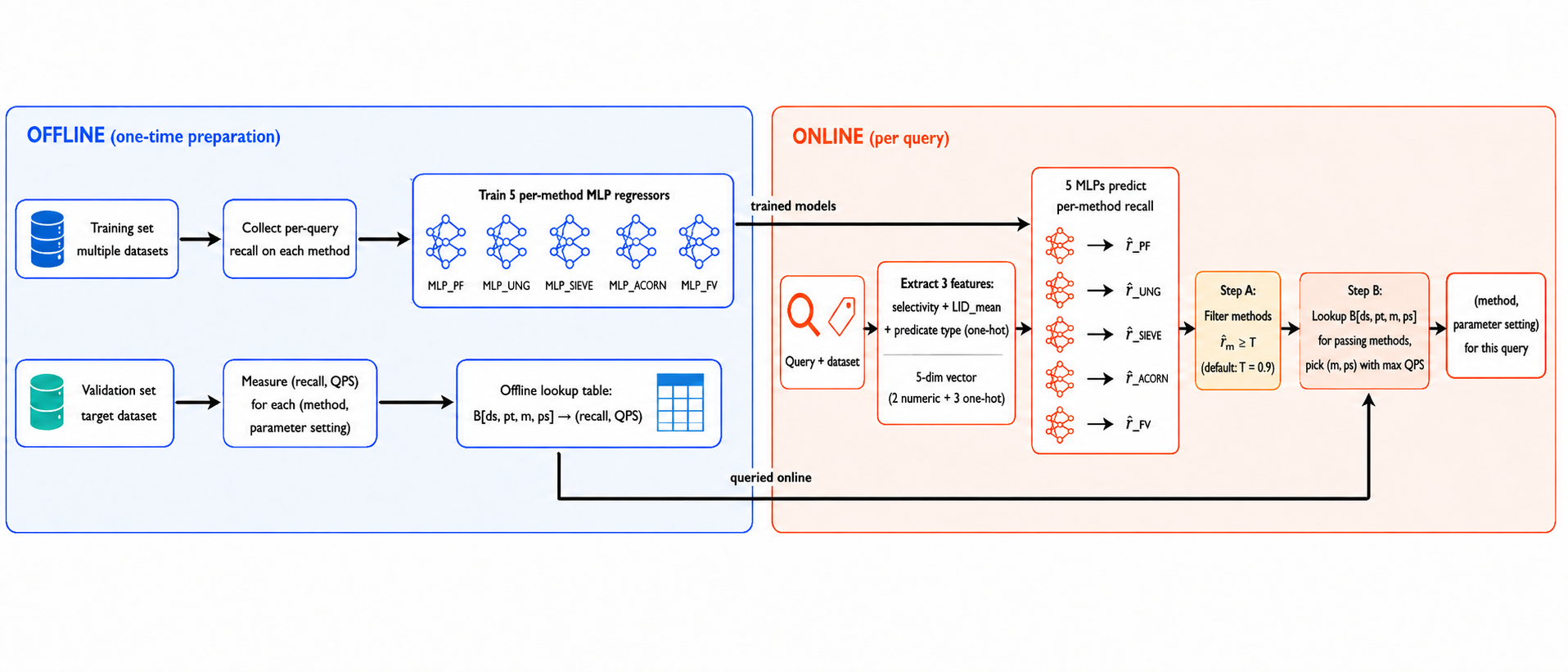}
    \captionsetup{justification=centering, singlelinecheck=false}
    \caption{Query-aware ML routing pipeline.
    }
    \label{fig:ml-router-arch}
\end{figure*}
Our query-aware ML router instantiates $\mathcal{R}$ as a two-stage pipeline as demonstrated in Figure~\ref{fig:ml-router-arch}.
The pipeline contains two main stages. For the offline stage, we train a model for each of the five candidate methods and build a lookup table $B$ on the validation datasets.
For the online stage, a query and its dataset are input. We first extract their features, which are fed to five trained models. Then we can get predicted recall for each candidate method, filter them by the predefined threshold $T$, search over the table $B$, and finally get a combination of method and parameter setting with maximum QPS.

Next, we will focus on feature selection and model design.

\subsection{Feature Selection}
\label{ssec:features}

We start with $22$ candidate features in three groups:
\begin{enumerate}
\item[(1)] $6$ query-aware features, i.e., number of labels in query, selectivity, minimum per-label frequency, maximum per-label frequency, mean per-label frequency, and label co-occurrence. 

Specifically, 

\begin{itemize}
\item \emph{selectivity} is the fraction of base vectors satisfying the predicate and defined as $\mathrm{sel}(q) = |\{(v_i, L_i) \in V : P(L_i, L_q) = \text{true}\}|/n$; 
\item \emph{label co-occurrence} is the fraction of base vectors whose label set contains all of $q$'s labels and denoted as $|\{v \in V: L_q \subseteq L_v\}|/n$.
\item the three \emph{per-label frequency statistics} are the minimum, maximum, and mean of $f(l_i) = |\{v \in V : l_i \in L_v\}|/n$ taken over the labels $l_i \in L_q$;
\end{itemize}

\item[(2)] $15$ dataset-level feature, i.e., dataset size, dimensionality, LID mean, LID median, LID standard deviation, relative-contrast~\cite{HeKumarChang2012RC} median, relative-contrast 5--95\% trimmed mean, relative-contrast 95th percentile, label cardinality, label entropy, number of unique label combinations, average labels per vector, distribution factor (mean sliced Wasserstein distance~\cite{Bonneel2015SlicedWasserstein}), correlation ratio, and normalized correlation ratio. 
\begin{itemize}
\item the three \emph{LID statistics} are the mean, median, and standard deviation of $\mathrm{LID}(q)$ over a fixed sample of base vectors; 
\item the three \emph{relative-contrast statistics} are the median, 5--95\% trimmed mean, and 95th percentile of $\mathrm{RC}(q)$ over the sample queries where $\mathrm{RC}(q) = \mathrm{dist}(q, k\text{-th NN}) / \mathrm{dist}(q, 1\text{st NN})$;
\item \emph{Label entropy} is $-\sum_{l \in U} p(l) \log p(l)$ with $p(l) = |\{v : l \in L_v\}|/n$; 
\item \emph{number of unique label combinations} is $|\{L_v : v \in V\}|$; 
\item \emph{average labels per vector} is $\frac{1}{n}\sum_{v \in V} |L_v|$;
\item \emph{distribution factor} is the mean sliced Wasserstein distance between each label's vector subset and the global base distribution;
\item \emph{correlation ratio} is the size-weighted mean of per-label LID divided by the global LID; 
\item \emph{normalized correlation ratio} rescales the correlation ratio by its expected value under a random subset of the same size, removing small-sample LID bias. 
\end{itemize}
\item[(3)] $1$ feature for predicate type.
\end{enumerate}

The first $21$ features are numeric and the last one is categorical.
Since using all features risks overfitting, we perform feature selection. We rank the numeric features by RandomForest feature importance on the training set, then construct nested subsets of different sizes in the order of importance, and train the ML model under five random seeds for each size to average out initialization noise. The full setup, results, and analysis appear in Section~\ref{sssec:feature-selection}.

Based on our experimental results (Section~\ref{sssec:feature-selection}), we finally select three features: selectivity, $\mathrm{LID}_{\mathrm{mean}}$, predicate type. They will be combined into one feature vector as the input of the ML model.

\subsection{Model Design}
\label{ssec:regression}

An ML model can usually be divided into classification and regression. We choose regression (which predicts each method's recall) over classification (which directly predicts the best method) because numeric recall output of different methods carries strictly more information than a single discrete label. It will keep the decision stable when several methods sit at similar recall levels. This choice can be validated in Section~\ref{sssec:cls-vs-reg}.
We chose the Multi-layer Perception regression model (\underline{MLP-Reg}) as our final choice.

For each candidate method $m \in \mathcal{M}$, we independently train a MLP regressor $f_m$. Its input is the feature vector $\mathbf{x}(q, \mathit{ds})$, and its output is the predicted recall@$10$ that $m$ would attain on query $q$. This design is flexible. As each candidate method uses its independent MLP regressor, it only requires training the corresponding new $f_m$ without retraining the existing ones when a new candidate method is added to the router. 
Each $f_m$ is a 2-layer MLP with hidden layer sizes \texttt{(64, 32)}, ReLU activation, MSE loss, and Adam optimization. This depth is selected by the layer ablation study in Section~\ref{sssec:mlp-depth} as a balance of filtered ANN search's recall and inference latency.

\begin{algorithm}[!t]
\caption{Per-Query ML Routing}
\label{alg:per-query-routing}
\begin{algorithmic}[1]
\REQUIRE query $q$, dataset context $\mathit{ds}$, deployment threshold $T$
\ENSURE (method $m^*$, parameter setting $\mathit{ps}^*$)
\STATE $\mathbf{x} \gets \mathrm{ExtractFeatures}(q, \mathit{ds})$ \hfill // Section~\ref{ssec:features}
\FOR{each $m \in \mathcal{M}$}
    \STATE $\hat{r}_m \gets f_m(\mathbf{x})$ \hfill // five MLP forwards
\ENDFOR
\STATE $\mathcal{P} \gets \{m \in \mathcal{M} : \hat{r}_m \ge T\}$ \hfill // $T$-threshold filter
\IF{$\mathcal{P} \neq \emptyset$}
    \FOR{each $m \in \mathcal{P}$}
        \STATE $\mathit{ps}_m \gets \arg\max_{\mathit{ps}} \mathrm{QPS}(B[\mathit{ds}, \mathit{pt}, m, \mathit{ps}])$ s.t. $\mathrm{recall}(B[\mathit{ds}, \mathit{pt}, m, \mathit{ps}]) \ge T$
    \ENDFOR
    \STATE $m^* \gets \arg\max_{m \in \mathcal{P}} \mathrm{QPS}(B[\mathit{ds}, \mathit{pt}, m, \mathit{ps}_m])$
    \STATE $\mathit{ps}^* \gets \mathit{ps}_{m^*}$
\ELSE
    \STATE $m^* \gets \arg\max_m \hat{r}_m$ \hfill // fallback
    \STATE $\mathit{ps}^* \gets \arg\max_{\mathit{ps}} \mathrm{QPS}(B[\mathit{ds}, \mathit{pt}, m^*, \mathit{ps}])$ s.t. $\mathrm{recall} \ge T$ (or the max-recall setting if none satisfies)
\ENDIF
\STATE \textbf{return} $(m^*, \mathit{ps}^*)$
\end{algorithmic}
\end{algorithm}

The final ML routing algorithm can be obtained by combining the regression models with the offline benchmark table $B$, as shown in Algorithm~\ref{alg:per-query-routing}.
Lines 1--4 extract features and predict per-method recall. Line 5 filters methods by the deployment threshold $T$, yielding the passing set $\mathcal{P}$. If $\mathcal{P}$ is non-empty (lines 6--11), the algorithm picks the (method, parameter setting) pair with the maximum QPS within $\mathcal{P}$: for each surviving method it first selects from $B$ the parameter setting whose mean recall on the current (dataset, predicate type) combination is at least $T$ and whose QPS is maximum; then it picks the method with the overall best QPS. If $\mathcal{P}$ is empty (lines 13--15, fallback), the algorithm picks the method with the highest predicted recall and pairs it with the max-recall parameter setting available, getting as close to the precision target as possible.

The threshold $T$ acts as a flexible and deployment-oriented knob: the same trained model can serve different values of $T$ (typically $0.5$ to $0.99$). A larger $T$ favors high recall at lower QPS; a smaller $T$ favors high QPS at lower recall; sweeping $T$ traces out a full recall--QPS Pareto curve from a single training run (see Section~\ref{sec:main-result} and Figure~\ref{fig:main-result}). The fallback branch occurs only under extreme thresholds, such as $T = 0.99$.

%% file: RelatedWork.tex
\section{Related Work}
\label{sec:relatedwork}

This paper is related to three research topics: filtered ANN methods (Section~\ref{sec:rw-fanns}), adaptive query planning and method selection (Section~\ref{sec:rw-routing}), and related benchmarks (Section~\ref{sec:rw-bench}).

\subsection{Filtered ANN Search}
\label{sec:rw-fanns}

Existing filtered ANN (FANNS) methods can be categorized into two types based on the filtered attribute type: \textit{categorical filter ANN} and \textit{range filtered ANN} methods.

Categorical Filtered ANN methods can be further divided into three types based on the execution strategy ~\cite{FilteredANNBenchmark2024}.
\textit{Filter-then-search} methods first apply the attribute constraint to obtain a candidate subset, then perform ANN search on this subset. Representative methods include Pre-filter, which performs a brute-force search on the filtered subset, and UNG~\cite{UNG2024}, which builds per-attribute sub-graph indexes connected by a label-navigating graph for cross-partition traversal. These methods perform well under high selectivity but degrade toward linear scan as selectivity drops.
\textit{Search-then-filter} methods first perform ANN search on a global index, then apply the attribute constraint to the retrieved candidates. Representative methods include Post-filter HNSW~\cite{HNSW2018} and Post-filter IVFPQ~\cite{FaissGPUIVFPQ2019}. While they directly reuse mature ANN index structures, under strong selectivity few of the initial top-$k'$ candidates satisfy the constraint; maintaining recall therefore requires substantial candidate-set expansion at significant efficiency cost.
\textit{Hybrid-search} methods embed attribute constraints directly into ANN index construction or search. HQANN~\cite{HQANN2022} introduces lightweight attribute-aware filtering on HNSW; Filtered-DiskANN and Stitched-DiskANN~\cite{FilteredStitchedVamana2023} apply label-aware neighbor selection on the Vamana graph; ACORN-1 and ACORN-$\gamma$~\cite{ACORN2024} extend HNSW with attribute-aware pruning and multi-hop neighbor expansion; SIEVE~\cite{SIEVE2025} leverages historical query workloads to offline-construct a heterogeneous collection of sub-indices targeting common filter patterns; CAPS~\cite{CAPS2023} and NHQ~\cite{NHQNPG2023} target fixed-length-label scenarios, with CAPS partitioning the attribute space via K-means and AFT, and NHQ fusing vector and attribute distances into a unified weighted distance. Although these methods generally achieve more balanced performance across filter scenarios, each retains a structural preference for particular filter types (containment, overlap, equality) and data distributions.

Range filtered ANN methods filters 
over numerical attributes such as timestamps and prices.
Representative methods include iRangeGraph~\cite{iRangeGraph2024} (range-dedicated graphs assembled from sub-graphs), SeRF~\cite{SeRF2024} (segment-graph augmentation of HNSW), ARKGraph~\cite{ARKGraph2023} (all-range $k$-NN graph), UNIFY~\cite{UNIFY2025} (unified index spanning the full range axis), WindowFilters~\cite{WindowFilters2024} (window-based precomputed indices), RangePQ~\cite{RangePQ2025} (efficient dynamic indexing for range queries), and TimestampANN~\cite{TimestampANN2025} (timestamp-specialized variant). As noted in the FANNS benchmark~\cite{FilteredANNBenchmark2024}, these methods are largely incompatible with the categorical-label algorithms discussed above and fall outside the scope of this paper.

\subsection{Adaptive Query Planning and Method Selection}
\label{sec:rw-routing}

Our routing work instantiates the \emph{algorithm selection} problem~\cite{Rice1976} in the FANNS domain. This paradigm has mature portfolio-based solutions in adjacent fields such as SAT solving~\cite{SATzilla2008}, where the optimal solver is predicted from problem-instance features rather than chosen as a single universal method. Similar ideas have only recently begun to extend to vector retrieval.
The most closely related work is the learning-based query planner of Gan and Wang~\cite{LearningBasedQueryPlanning2026}, which trains a per-dataset two-layer MLP classifier to select between Pre-filter and Post-filter on a per-query basis, using lightweight features (selectivity, dimensionality, dataset distribution). 
This work reports up to $4\times$ speedup over single-strategy baselines. However, its strategy space is restricted to these two basic methods, excluding hybrid-search approaches such as UNG, ACORN, and SIEVE. Moreover, as the planner is trained on each dataset, no empirical evidence of cross-dataset generalization is reported.

A complementary line of work comes from unfiltered vector similarity search (VSS). Iceberg~\cite{Iceberg} addresses method selection for general VSS from a task-centric view, proposing the Information Loss Funnel model and deriving an interpretable decision tree over easy-to-compute meta-features such as the Davies--Bouldin index (DBI) for clustering tightness, vector-norm coefficient of variation (CV), relative angle (RA), and relative contrast (RC), to guide selection among methods such as HNSW, NSG, and RaBitQ. While Iceberg does not address attribute filtering, its methodology of \emph{constructing an interpretable decision tree from dataset meta-features} is structurally analogous to ours, validating meta-feature routing as a viable direction in the broader VSS area.

Our work differs from the above in three notable ways: (1)~\emph{broader method coverage} --- the candidate pool spans all three FANNS execution paradigms (filter-then-search, search-then-filter, hybrid-search) rather than only Pre- vs.\ Post-filter; (2)~\emph{both rule-based and learned routers are provided} --- the former is derived directly from structural analysis via a three-feature decision tree (query scenario, label cardinality, LID), while the latter approximates the Oracle upper bound via per-method regression modeling; (3)~\emph{systematic evaluation of generalization} --- we evaluate on five mid-scale out-of-sample validation datasets (500K--800K vectors, including Yahoo and DBpedia real-text data), directly testing routing transferability beyond the training distribution.

\subsection{Benchmarks for Filtered ANN}
\label{sec:rw-bench}

The most directly related benchmark is the unified FANNS benchmark of Shi et al.~\cite{FilteredANNBenchmark2024}, which systematically evaluates 10 FANNS algorithms (NHQ, Filtered-DiskANN, Stitched-DiskANN, ACORN-1, ACORN-$\gamma$, CAPS, UNG, Pre-filter Brute-Force, Post-filter HNSW, Post-filter IVFPQ) on 6 real-world datasets (including YFCC and YouTube) under five filter modes (containment, overlap, equality, fixed-equality, combined). It categorizes methods into the filter-then-search, search-then-filter, and hybrid-search paradigms, and emphasizes parameter fairness in evaluation. Our empirical foundation builds directly on this benchmark: on top of its method and dataset pool, we introduce dataset difficulty features (LID, RC, distribution factor) as routing signals, and independently construct five mid-scale validation datasets (including Yahoo and DBpedia real-text data) to assess routing generalization beyond the training set.
General ANN benchmarks such as ANN-Benchmarks~\cite{ANNBenchmarks2020} and Big-ANN-Benchmarks~\cite{simhadri2022bigann} are mature in the unfiltered setting, but their evaluation protocols do not cover attribute-filtered scenarios. Iceberg~\cite{Iceberg} extends end-to-end task-centric evaluation in the general VSS setting but likewise does not address filter constraints. Our work complements this ecosystem by transitioning from benchmarking toward \emph{method selection grounded in structural observation} in the FANNS setting.

%% file: Experiment.tex
\section{Experiments}
\label{sec:experi}

All experiments run on a Linux node equipped with an AMD EPYC 9654 processor (96 cores, 192 threads, 594\,GB RAM). For each experiment, 16 CPU cores are allocated with 128\,GB of memory; multi-threaded runs use 16 threads. 

\subsection{Experimental Setup}
\label{sec:setup}

\subsubsection{Methods}
\label{sec:method-list}
Two batches of method are used in different stages of the experimental study: 

\begin{itemize}
    \item The first batch is ten filtered ANN methods used in our initial benchmark, which are used to implement extensive experimental study and derive our motivation to propose novel router. They are Pre-filter \cite{FilteredANNBenchmark2024}, Post-filter \cite{FilteredANNBenchmark2024}, UNG\cite{UNG2024}, SIEVE\cite{SIEVE2025}, ACORN-$\gamma$, ACORN-1\cite{ACORN2024}, NHQ\cite{NHQNPG2023}, CAPS\cite{CAPS2023}, FilteredVamana and StichedVamana\cite{FilteredStitchedVamana2023}, which cover almost all existing categorical filtered ANN methods. Our systematic experimental results for these methods are shown in Figure~\ref{fig:methods-comparison}.

    \item The second batch is the baseline methods, which will be compared with our proposed methods. We select five methods (UNG, Post-filter, SIEVE, ACORN-$\gamma$, and FilteredVamana) from the first batch, as they have optimal performance on at least one combination of dataset and predicate type, as shown in Figure~\ref{fig:methods-comparison}. 
The remaining methods in the empirial study are excluded from the candidate set out of different reasons: Pre-filter has recall = 1, however its QPS is far lower than the other methods%
; ACORN-1 is the $\gamma=1$ special case of ACORN-$\gamma$, and ACORN-1 either performs comparably to ACORN-$\gamma$ or is dominated by it; CAPS and NHQ restrict labels to fixed-length formats and cannot handle the variable-length label structures present in real world depolyment. We further observed that StitchedVamana exhibits stability issues across many parameter parameter settings: some parameter settings crash with segmentation faults, while others
produce corrupted indices that yield zreo recall.

We also include our RuleRouter (introduced in section \ref{sec:method1}) as one baseline method, which is a hand-crafted router that select the best performed method based on dataset-level features and predicate type.

Moreover, Gan~\cite{LearningBasedQueryPlanning2026} proposes a learned binary planner that picks either Pre-filter or Post-filter execution for each query. Since both pre-filter and Post-filter are already sit within our setting, the action space of \cite{LearningBasedQueryPlanning2026} is therefore a strict subset of ours. So we do not include this work as our baseline.

\end{itemize}

We report recall@10 and average QPS over 16 threads as the primary metrics; the ML router's QPS aggregates both the routing time and per-query search latency.

\subsubsection{Training}

\paragraph{Datasets.}
We use six real-world training datasets~\cite{FilteredANNBenchmark2024} as summarised in Table~\ref{tab:training-datasets}. These datasets span a wide range of domain, size, dimensionality, and label cardinality, along with $\mathrm{LID}_{\mathrm{mean}}$ and $card(V)$ as shown in Table \ref{tab:best-method-per-cell}. They cover the typical workload characteristics of filtered ANN retrieval and can be downloaded\footnote{\label{fn:datasets}\url{https://huggingface.co/datasets/ffa500/filterbenchmark}}.

\begin{table}[!htbp]
\centering
\caption{Six real-world training datasets.}
\label{tab:training-datasets}
\small
\setlength{\tabcolsep}{4pt}
\begin{tabular}{@{}llrrr@{}}
\toprule
Dataset & Domain & Size & Dim & \#Labels \\
\midrule
arxiv      & academic paper embeddings   & 132K & 768  & 4{,}231 \\
yfcc       & Flickr images (YFCC100M)    & 1M   & 192  & 181{,}931 \\
LAION-1M   & image--text pairs           & 1M   & 512  & 30 \\
tripclick  & travel-search logs          & 1M   & 768  & 29 \\
ytb\_audio & YouTube-8M audio            & 5M   & 128  & 3{,}862 \\
ytb\_video & YouTube-8M video            & 1M   & 1024 & 3{,}862 \\
\bottomrule
\end{tabular}
\end{table}

\paragraph{Training data collection.}
\label{ssec:training-data}
For each training data containing the combination of training dataset, predicate type, and candidate methods, we first perform a parameter sweep over the method's parameter space. Candidate methods are five methods in the second batch of method as mentioned in section \ref{sec:method-list}. Table~\ref{tab:sweep-ranges} lists each method's sweep range and a typical best parameter setting as a reference.
Then we select the parameter setting with the best recall--QPS tradeoff for each training data and run the full query workload under the selected parameter setting to record per-query recall@10. 

The resulting training set contains approximately $6 \times 3 \times 1{,}000 \times 5 \approx 90{,}000$ records, which are produced under 6 training datasets, 3 predicate types, $1{,}000$ queries on each dataset, 5 candidate methods. The candidate methods serve as labels for the MLP-Reg model.

\begin{table*}[!htbp]
\centering
\captionsetup{justification=centering, singlelinecheck=false}
\caption{Parameter sweep ranges for training data collection.
}
\label{tab:sweep-ranges}
\small
\setlength{\tabcolsep}{4pt}
\begin{tabularx}{\textwidth}{l X X X}
\toprule
Method & Build parameters & Search parameter & Typical best config \\
\midrule
UNG            & max\_degree=\{32,48,64,96\}, $L_\text{build}$=\{100,150,200\} & $L_\text{search}$=\{100,300,500\} & max\_degree=96, $L_\text{build}$=200, $L_\text{search}$=500 \\
\midrule
Post-filter    & $M$=\{32,48,64\}, efc=\{100,200,400\} & ef=\{1200,1500,2000\} & $M$=64, efc=100, ef=2000 \\
\midrule
SIEVE          & $M$=\{16,32\}, index\_budget=\{1,2,3\}, hist\_pct=\{0.25,0.5\} & ef\_search=[30,200] & $M$=32, index\_budget=2.0, hist\_pct=0.25, ef\_search=200 \\
\midrule
ACORN-$\gamma$ & $M$=\{48,64\}, $M_\beta$=\{48,64,96\}, $\gamma$=\{1,4,8,12,24\} & ef=\{1000,1200\} & $M$=64, $M_\beta$=96, $\gamma$=8, ef=1200 \\
\midrule
FilteredVamana & $R$=\{32,64,128\} & $L_\text{search}$=\{100,200,500,1000,2000\} & $R$=128, $L_\text{search}$=2000 \\
\bottomrule
\end{tabularx}
\end{table*}

We select the best parameter setting for each candidate method under all combinations of dataset and predicate type, rather than a globally fixed setting. This ensures that the training labels reflect each method's potential best performance for that specific combination. It can avoid the underestimation that would arise if a parameter setting is globally reasonable but locally underfit. The drawback is that method selection and parameter setting selection must be modeled separately. Specifically, MLP-Reg predicts the query performance of each candidate method given one query.

\subsubsection{Validation}

\paragraph{Datasets.}
There are five validation datasets that router never sees during training, as summarised in Table~\ref{tab:validation-datasets}. Their dimensions and label cardinality generally follow that of the training datasets. Three synthetic datasets are generated by sampling Zipf-distributed label sets over Gaussian vector clusters with a fixed seed.
The two real-world text datasets (yahoo800k and dbpedia560k), both encoded with the same sentence-transformer model, evaluate the router on natural-language workloads that differ structurally from distributions of the training datasets. For each combination of dataset and predicate type, we sample $1{,}000$ queries, yielding $15{,}000$ validation queries in total.

\begin{table}[!htbp]
\centering
\caption{Five validation datasets.%
}
\label{tab:validation-datasets}
\small
\setlength{\tabcolsep}{4pt}
\begin{tabular}{@{}llrrr@{}}
\toprule
Dataset & Domain & Size & Dim & \#Labels \\
\midrule
synth\_192d     & synthetic (Zipf/Gaussian)                    & 800K & 192  &   200 \\
synth\_512d     & synthetic (Zipf/Gaussian)                    & 800K & 512  &    30 \\
synth\_768d\_hc & synthetic (Zipf/Gaussian)                    & 800K & 768  & 1{,}000 \\
\midrule
yahoo800k       & Yahoo Answers topics~\cite{Zhang2015CharCNN} & 800K & 768  &    14 \\
dbpedia560k     & DBpedia ontology~\cite{Zhang2015CharCNN}     & 560K & 768  &    14 \\
\bottomrule
\end{tabular}
\end{table}

\paragraph{Dataset preparation and query generation.}
We use each dataset as it ships from its source. Each dataset comes with two parts: a \emph{base set} (the vectors over which the index is built) and a separate \emph{query set} (a small subset of vectors held out for testing); we use both as released. Every base vector $v_i$ carries a label set $L_i$ from the dataset's metadata (subject classifications for arxiv, image tags for yfcc, $30$ categorical labels for LAION1M).

To evaluate the router under arbitrary query workloads, we generate $1{,}000$ queries for each combination of dataset and predicate type. 
Each query has two parts: a query vector $q$ and a query label set $L_q$. The query vectors are drawn from a pre-defined set that is disjoint from the dataset. 
Specifically, for the two real-world text datasets (yahoo800k and dbpedia560k), the query set originates from the source dataset; for the three synthetic datasets, query vectors are constructed by adding Gaussian noise (scale = 10\% of median base-vector norm) to randomly chosen vectors from the datasets.
We choose the size of $L_q$ to match how real queries are typically issued: for Equality and AND predicate types, queries are usually specified with a few labels, so $|L_q|$ is small (1--3); for OR predicate type, query labels are typically span a wide range, so $|L_q|$ is broader.

The ground-truth results for validation
are obtained through brute-force search over all candidate vectors $v_i$ whose label set $L_i$ satisfies the query predicate with $L_q$.

\subsection{Design Ablations}
\label{sec:compared}

In this section, we conduct ablation studies for the ML router from three perspectives: feature selection, model type (classification vs.\ regression), and MLP depth.

\paragraph{(a) Feature Selection.}
\label{sssec:feature-selection}

Starting from the $22$ candidate features described in Section~\ref{ssec:features}, we will test which are actually needed by MLP-Reg.
We rank numeric candidate features by RandomForest feature importance on the training set and construct a nested family of subsets with different sizes, each prepended with \texttt{predicate type}.
We use $n$ to denote the number of most important features.
For every $n$, we train MLP-Reg under five random seeds
and report the mean and standard deviation of validation recall@10.
Multi-seed averaging is necessary because MLP training is sensitive to initialisation, and a single-seed estimate at the same $n$ would make the curve essentially unreadable.

\begin{figure}[!htbp]
\centering
\includegraphics[width=0.7\linewidth]{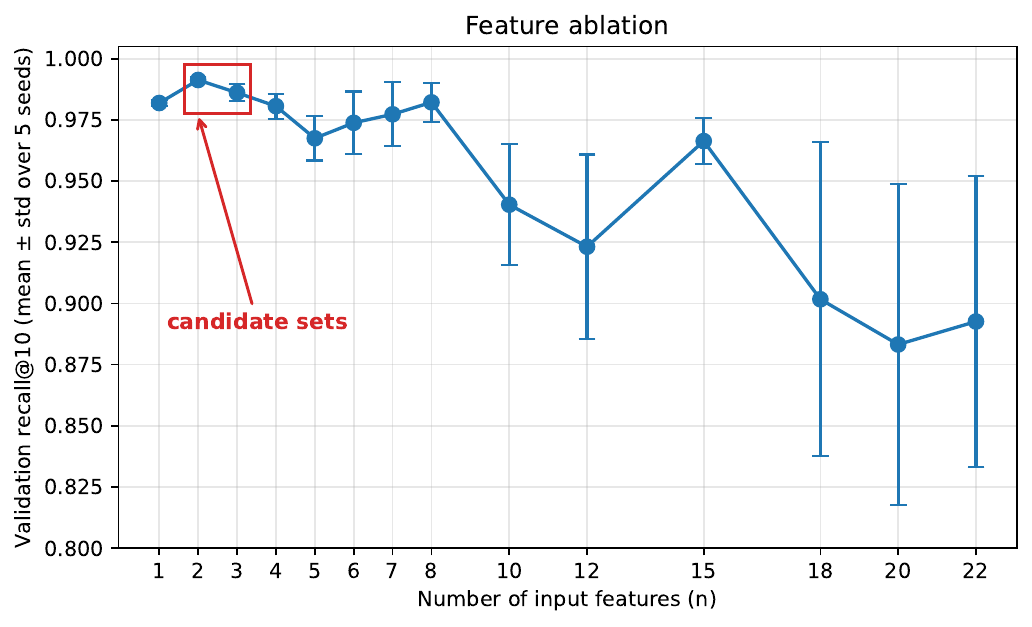}
\captionsetup{justification=centering, singlelinecheck=false}
\caption{MLP-Reg validation recall vs.\ feature count.%
}
\label{fig:ablation-features}
\end{figure}

The validation recall under different feature numbers is shown in Figure~\ref{fig:ablation-features}.
For $n \ge 10$ the curve becomes non-monotone and per-seed variance grows by an order of magnitude (std up to $0.06$). This can be explained that individual feature additions can either lift or depress mean recall depending on whether the added feature is informative or merely a dataset ``fingerprint'' that overfits the six training dataset. We therefore restrict the number of candidate feature to $n \le 8$, where the mean recall is stable and high across seeds, and pick candidates in this range.
Within the $n \le 8$ plateau, the model achieves the highest recall with $n = 2$ and $n = 3$.
The two candidates are separated by only $0.005$ recall and both have tight per-seed variance, so recall alone does not distinguish them. We therefore evaluate them on per-query latency before committing to the final minimal feature set.
Table~\ref{tab:feature-latency-tradeoff} reports the per-query latency on two real-world validation datasets.
On both real-world text datasets, $n = 3$ is $1.7$ to $5.6$ times faster than $n = 2$, because the \texttt{LID\_mean} feature steers the router away from latency-heavy methods such as UNG on AND/OR queries (where UNG must scan many label partitions per query, which is slow on the high-LID text embeddings these datasets produce). So we adopt $n=3$ as our final number of features.

\begin{table}[!htbp]
\centering
\captionsetup{justification=centering, singlelinecheck=false}
\caption{Per-query latency under two candidate feature sets
}
\label{tab:feature-latency-tradeoff}
\small
\begin{tabular}{@{}lrr@{}}
\toprule
Dataset             & $n=2$ latency ($\mu$s) & $n=3$ latency ($\mu$s) \\
\midrule
dbpedia560k         & $8559$                 & $\mathbf{4993}$~~~($1.7\times$) \\
yahoo800k           & $20974$                & $\mathbf{3727}$~~~($5.6\times$) \\
\bottomrule
\end{tabular}
\end{table}

\paragraph{(b) Classification vs.\ Regression.}
\label{sssec:cls-vs-reg}

The ML model can be trained either as a classifier or as a regressor.
The two model families share the same input features, training set, and network capacity, but they differ in training objective and output form. Specifically, classification directly outputs the discrete top-1 method label, whereas regression outputs a continuous predicted recall for each candidate method.

For a fair comparison of the two training objectives, we reduce the regression output to plain argmax (selecting the method with the highest predicted recall) to align with the classifier's top-1 output.
Note that the model in the deployment evaluation of Section~\ref{sec:main-result} additionally filters methods by their predicted recall ($\hat{r}_m \geq T$) before selecting the lookup-table max-QPS parameter setting, which a classifier cannot do since it emits only a single label.
The plain argmax evaluation in this section therefore underestimates regression's deployment time performance.

\begin{table}[!htbp]
\centering
\captionsetup{justification=centering, singlelinecheck=false}
\caption{Recall@10 of classification and regression models
}
\label{tab:cls-vs-reg}
\small
\begin{tabular}{@{}llccc@{}}
\toprule
Family & Router & yahoo800k & dbpedia560k & Aggregate \\
\midrule
\multirow{3}{*}{Classification}
 & LogisticReg  & 0.892 & 0.931 & 0.903 \\
 & MLP          & 0.871 & 0.933 & 0.937 \\
 & RandomForest & 0.940 & 0.927 & 0.958 \\
\midrule
\multirow{3}{*}{Regression}
 & Ridge        & 0.954 & 0.977 & 0.985 \\
 & MLP-Reg      & 0.957 & 0.993 & \textbf{0.986} \\
 & RF-Reg       & 0.950 & 0.994 & \textbf{0.987} \\
\bottomrule
\end{tabular}
\end{table}

We select three representative algorithms from each model family: classification uses LogisticRegression, MLP, and RandomForest; regression uses Ridge, MLP-Reg, and RF-Reg.
Table~\ref{tab:cls-vs-reg} demonstrates their recall in two datasets (yahoo800k, dbpedia560k) along with the aggregate recall (denoted as Aggregate in the Table) on all five validation datasets.
The regression family is significantly better than the classification family overall.
The gap stems from how much information the two training losses can exploit. Classification is trained with cross-entropy loss, which only checks whether the argmax falls on the ground-truth method: a prediction is either correct or wrong, with no distinction between being slightly off (picking a method whose recall is $0.005$ below the best) and being far off (picking one $0.5$ below). On validation set where almost every method's recall sits close to the ceiling, ties are frequent and the cross-entropy gradient cannot smoothly steer the model toward the truly optimal method. Regression instead is trained with MSE on each candidate's measured recall, giving a continuous, differentiable error that grows with the distance between predicted and true recall. The gradient is proportional to the induced recall loss, so the argmax stays stable when several methods sit at similar recall.
We therefore choose regression as the training objective.

Within the regression family, the three models have nearly identical overall recall (Ridge $0.985$ / MLP-Reg $0.986$ / RF-Reg $0.987$), but inference latency differs by an order of magnitude ($0.13$ / $0.74$ / $7.83\,\mu$s per query): RF-Reg is $10.5\times$ slower than MLP-Reg.
Filtered ANN queries in high-QPS workloads commonly take hundreds of microseconds, so any routing overhead exceeding $1$--$2\,\mu$s starts to reduce the overall QPS.
We therefore select MLP-Reg, since its recall is only $0.001$ below the strongest RF-Reg and its latency is an order of magnitude lower.

\paragraph{(c) MLP Depth.}
\label{sssec:mlp-depth}

Since the input is only $3$ features, the model does not need to be deep. We adopt a $2$-hidden-layer MLP $(64, 32)$: with so few inputs, a single hidden layer has limited capacity to capture interactions among the features (in the worst case collapsing toward a logistic-regression-like decision boundary), while two layers reach adequate approximation capacity at sub-microsecond inference cost. Deeper hidden layers (such as $3$-$4$) bring no obvious recall gain on this low-dimensional input but raise inference latency $3$-$7\times$ per query as shown in Table~\ref{tab:mlp-depth}.

\begin{table}[t]
  \centering
  \caption{MLP depth ablation: routing recall and inference latency.}
  \label{tab:mlp-depth}
  \begin{tabular}{ccc}
    \toprule
    \#Layers & Recall & $\mu$s/query \\
    \midrule
    2 & $0.9863$ & $0.50$ \\
    3 & $0.9896$ & $1.50$ \\
    4 & $0.9874$ & $3.38$ \\
    \bottomrule
  \end{tabular}
\end{table}

\subsection{Main Result: Full Recall-QPS Comparison}
\label{sec:main-result}

\begin{figure*}[!tbp]
\centering
\includegraphics[width=\textwidth]{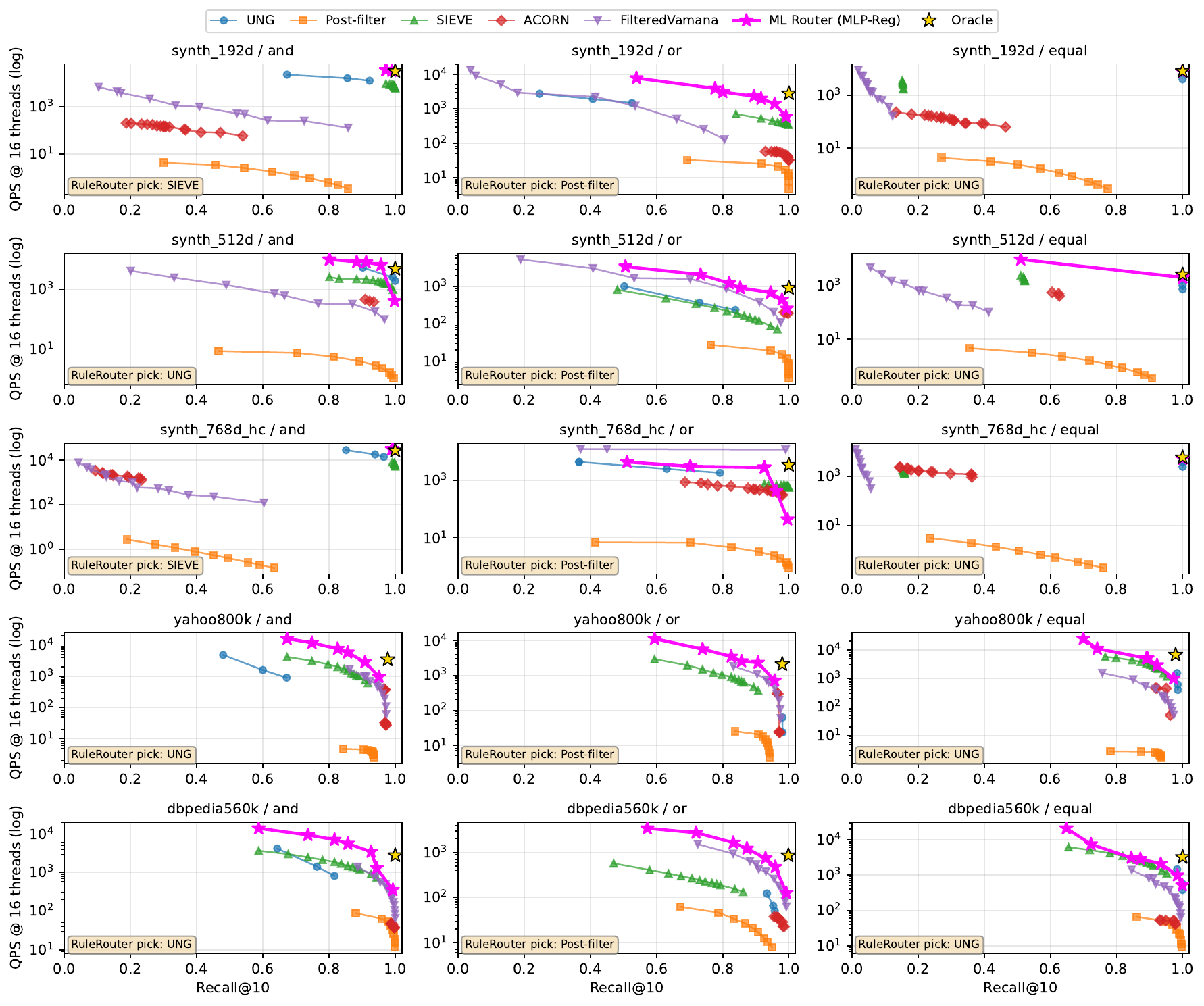}
\caption{Recall--QPS Pareto on all combinations of dataset and predicate type.}
\label{fig:main-result}
\end{figure*}

Having fixed the ML Router design in section \ref{sec:compared}, we compare the router against all baselines and RuleRouter in the (recall, QPS) plane on all five validation datasets over three predicate types, with results shown in Figure~\ref{fig:main-result}.

\paragraph{Overall performance.}
We define the \textit{Oracle} as a hypothetical router that selects the method achieving the highest recall for each query, i.e., the \textit{Oracle} achieves the theoretical best performance.
The ML Router curve (magenta) has the best performance on almost all datasets and predicate types.
By routing each query to the method the model deems most suitable via per-query prediction, the ML Router achieves a multi-workload aggregate performance that no single method can match.

\paragraph{Gains over RuleRouter.}
Since RuleRouter picks one candidate method for each combination of validation dataset and predicate type, its performance will overlap with one baseline method. So we only mark its picked method rather than show its recall-QPS curve~\ref{fig:main-result}.
The ML Router's gains over RuleRouter manifest in two situations.
First, for the OR predicate type on several datasets (synth\_768d\_hc, yahoo800k, synth\_192d, dbpedia560k), the RuleRouter chooses Post-filter via an LID threshold, but Post-filter must use a very slow parameter setting to reach recall $\geq 0.9$ on these workloads, with QPS pushed down to the $10$--$100$ range.
The ML Router routes most queries to SIEVE or FilteredVamana, achieving overall QPS one to two orders of magnitude higher.
Second, for the AND predicate type on two datasets (yahoo800k, dbpedia560k), the rule chooses UNG, which attains a maximum mean recall of only $0.67$ and $0.82$ respectively across all sweep parameter settings, short of threshold $T$ = 0.9. Therefore, RuleRouter could fail to meet the recall threshold, which degrades its performance.

\paragraph{ML Router latency.}
\label{ssec:routing-latency}
To put the routing overhead in context, we measured the full streaming routing pipeline (Roaring-bitmap selectivity + feature scaling + 5 MLP-Reg forwards + offline config lookup) over all $1{,}000$ queries on all combinations of validation datasets and predicate types.
The 5 MLP forwards account for 41\,$\mu$s (median); Roaring-bitmap selectivity takes $\sim 0.4\,\mu$s on Equality (hash lookup over the precomputed set-count table) and 5--58\,$\mu$s on AND/OR (bitmap intersection/union over query labels). Across all $15{,}000$ queries, the routing per-query latency has median 54\,$\mu$s, p95 93\,$\mu$s, and max 167\,$\mu$s. The routing-to-query latency ratio is overall 0.2\% with a worst-case of 2.6\% on synth\_512d/AND.
Therefore, the routing latency is negligible compared to the query latency.

%% file: Conclusion.tex
\section{Conclusion}
\label{sec:conclu}
We presented a per-query ML routing framework for categorical filtered ANN query processing. Motivated by two key observations obtained from extensive experimental study, we propose a framework to train a lightweight router to predict each candidate method's recall on the incoming query, and selects the best method along with the parameter settting.
On five validation datasets unseen during training, the router attains average recall@10 $=$ 0.986, with 0.9\% gap to the ground truth.
Routing overhead has a median latency of around $54\,\mu$s per query, which is two orders of magnitude lower than the millisecond-scale latency of the underlying filtered ANN search. This demonstrates the practicality of the framework in real-world deployment.
Our two observations could also apply for range filtered ANN methods. Therefore, in the future, we plan to extend our framework to per-query routing among range filtered ANN methods.%